\documentclass{pasj02}

\usepackage{ulem,multicol}
\usepackage[whole]{bxcjkjatype}

\usepackage[switch,mathlines]{lineno}

\Received{$\langle$reception date$\rangle$}
\Accepted{$\langle$acception date$\rangle$}
\Published{$\langle$publication date$\rangle$}

\usepackage{graphicx}
\usepackage{amsmath}

\begin{document}

\title{
\LETTERLABEL
NinjaSat monitoring of Type-I X-ray bursts from the clocked burster SRGA J144459.2$-$604207}


\author{Tomoshi \textsc{Takeda}\altaffilmark{1,2}}
\author{Toru \textsc{Tamagawa},\altaffilmark{2,3,1}}
\author{Teruaki \textsc{Enoto},\altaffilmark{4,2}}
\author{Takao \textsc{Kitaguchi},\altaffilmark{2}}
\author{Yo \textsc{Kato}\altaffilmark{2}}
\author{Tatehiro \textsc{Mihara}\altaffilmark{2}}
\author{Wataru \textsc{Iwakiri}\altaffilmark{5}}
\author{Masaki \textsc{Numazawa}\altaffilmark{6}}

\author{Naoyuki \textsc{Ota}\altaffilmark{1,3}}
\author{Sota \textsc{Watanabe}\altaffilmark{1,2}}
\author{Arata \textsc{Jujo}\altaffilmark{1,3}}
\author{Amira \textsc{Aoyama}\altaffilmark{1,2}}
\author{Satoko \textsc{Iwata}\altaffilmark{1,3}}
\author{Takuya \textsc{Takahashi}\altaffilmark{1,3}}
\author{Kaede \textsc{Yamasaki}\altaffilmark{1,3}}

\author{Chin-Ping \textsc{Hu}\altaffilmark{7}}
\author{Hiromitsu \textsc{Takahashi}\altaffilmark{8}}

\author{Akira \textsc{Dohi}\altaffilmark{2,9}}
\author{Nobuya \textsc{Nishimura}\altaffilmark{10,2,11}}
\author{Ryosuke \textsc{Hirai}\altaffilmark{2,12,13}}

\author{Yuto \textsc{Yoshida}\altaffilmark{1,3}}
\author{Hiroki \textsc{Sato}\altaffilmark{14,3}}
\author{Syoki \textsc{Hayashi}\altaffilmark{1,3}}
\author{Yuanhui \textsc{Zhou}\altaffilmark{1,3}}
\author{Keisuke \textsc{Uchiyama}\altaffilmark{1,3}}
\author{Hirokazu \textsc{Odaka}\altaffilmark{15}}
\author{Tsubasa \textsc{Tamba}\altaffilmark{16}}
\author{Kentaro \textsc{Taniguchi}\altaffilmark{2}}




\altaffiltext{1}{Department of Physics, Tokyo University of Science, 1-3 Kagurazaka, Shinjuku, Tokyo 162-8601, Japan}
\altaffiltext{2}{RIKEN Cluster for Pioneering Research (CPR), 2-1 Hirosawa, Wako-shi, Saitama 351-0198, Japan}
\altaffiltext{3}{RIKEN Nishina Center, 2-1 Hirosawa, Wako, Saitama 351-0198, Japan}
\altaffiltext{4}{Department of Physics, Kyoto University, Kitashirakawa Oiwake, Sakyo, Kyoto 606-8502, Japan}
\altaffiltext{5}{International Center for Hadron Astrophysics, Chiba University, 1-33 Yayoi, Inage, Chiba, Chiba 263-8522, Japan}
\altaffiltext{6}{Department of Physics, Tokyo Metropolitan University, 1-1 Minamiosawa, Hachioji, Tokyo 192-0397, Japan}
\altaffiltext{7}{Department of Physics, National Changhua University of Education, Changhua, Changhua 50007, Taiwan}
\altaffiltext{8}{Department of Physics, Hiroshima University, 1-3-1 Kagamiyama, Higashi-Hiroshima, Hiroshima 739-8526, Japan}
\altaffiltext{9}{Interdisciplinary Theoretical and Mathematical Sciences Program (iTHEMS), RIKEN, 2-1 Hirosawa, Wako, Saitama 351-0198, Japan}
\altaffiltext{10}{Center for Nuclear Study (CNS), The University of Tokyo, 7-3-1 Hongo, Bunkyo, Tokyo 113-0033, Japan}
\altaffiltext{11}{National Astronomical Observatory of Japan (NAOJ), 2-21-1 Osawa, Mitaka, Tokyo 181-8588, Japan}
\altaffiltext{12}{School of Physics and Astronomy, Monash University, Clayton, VIC 3800, Australia}
\altaffiltext{13}{OzGrav: The ARC Centre of Excellence for Gravitational Wave Discovery, Clayton, VIC 3800, Australia}
\altaffiltext{14}{Department of System Engineering and Science, Shibaura Institute of Technology, 307 Fukasaku, Minuma, Saitama, Saitama 337-8570, Japan}
\altaffiltext{15}{Department of Earth and Space Science, Osaka University, 1-1 Machikaneyama, Toyonaka, Osaka 560-0043, Japan}
\altaffiltext{16}{Institute of Space and Astronautical Science, JAXA, 3-1-1 Yoshinodai, Chuo, Sagamihara, Kanagawa 252-5210, Japan}

\email{tomoshi.takeda@a.riken.jp}


\maketitle

\begin{abstract}
The CubeSat X-ray observatory NinjaSat was launched on 2023 November 11 and has provided opportunities for agile and flexible monitoring of bright X-ray sources. On 2024 February 23, the NinjaSat team started long-term observation of the new X-ray source SRGA J144459.2$-$604207 as the first scientific target, which was discovered on 2024 February 21 and recognized as the sixth clocked X-ray burster. Our 25-day observation covered almost the entire decay of this outburst from two days after the peak at $\sim$100~mCrab on February 23 until March 18 at a few~mCrab level. The Gas Multiplier Counter onboard NinjaSat successfully detected 12 Type-I X-ray bursts with a typical burst duration of $\sim$20 s, shorter than other clocked burster systems. As the persistent X-ray emission declined by a factor of five, X-ray bursts showed a notable change in its morphology: the rise time became shorter from 4.4(7) s to 0.3(3) s (1$\sigma$ errors), and the peak amplitude increased by 44\%. The burst recurrence time $\Delta t_{\rm rec}$ also became longer from 2~hr to 10~hr, following the relation of $\Delta t_{\rm rec} \propto F_{\rm per}^{-0.84}$, where $F_{\rm per}$ is the persistent X-ray flux, by applying a Markov chain Monte Carlo method. The short duration of bursts is explained by the He-enhanced composition of accretion matter and the relation between $\Delta t_{\textrm{rec}}$ and $F_{\rm per}$ by a massive neutron star. This study demonstrated that CubeSat pointing observations can provide valuable astronomical X-ray data.

\end{abstract}


\clearpage 


\section{Introduction}
\label{intro}



Type-I X-ray bursts (hereafter, X-ray bursts) are explosive transients in low-mass X-ray binaries (LMXBs), characterized by recurrent rapid increases in luminosity by order of magnitude, with emission continuing for several seconds to hours. 
These energetic bursts are triggered by nuclear burning in the accreting layer of a neutron star (NS) (for a review, see \cite{Galloway&Keek2021}). 
Key observables of X-ray bursts, such as rise time, peak flux, duration, and recurrence time, depend on the properties of the accreted matter, particularly the mass accretion rate and composition (e.g., \cite{Galloway2008}). 
Among over 115 known X-ray bursters \citep{Galloway2020}, only six sources exhibit notably regular burst recurrence times, called {\it clocked} bursters. GS~1826$-$24, which was first identified by Ginga X-ray satellite \citep{Makino1988}, is a prominent example of clocked bursters of which the periodic features were observationally confirmed by BeppoSAX and RXTE (e.g., \cite{Ubertini1999, Galloway2004}).

Clocked bursters are valuable for constraining theoretical models, as well as understanding the properties of the NS and companion star (\cite{2007ApJ...671L.141H, 2018ApJ...860..147M, 2020PTEP.2020c3E02D, 2021ApJ...923...64D, 2021PhRvL.127q2701H, 2022ApJ...929...72L,2022ApJ...937..124D}). 
Due to their regular burst recurrence times, one can investigate the relationship between the mass accretion rate and the physical conditions on the NS surface. The relation between the burst recurrence time $\Delta t_{\rm rec}$ and the persistent flux $F_{\rm per}$ with a constant $C$:
\begin{equation}
    \Delta t_{\rm rec} = C F_{\rm per}^{-\eta}
    \label{eq:eta}
\end{equation}
serves as a powerful diagnostic.
Assuming a critical fuel mass for burst ignition, a power-law index $\eta = 1$ is derived, where the persistent flux $F_{\rm per}$ is proportional to the mass accretion rate. Previous studies have found $\eta$ values ranging from approximately 1 to values greater than 1.\footnote{For instance, $\eta = 1.05\pm{0.02}$ for GS 1826$-$24 \citep{Galloway2004}, $\eta = 0.95 \pm{0.03}$ for MXB 1730$-$335 \citep{Bagnoli2013}, $\eta \gtrsim 1$ for IGR J17480$-$2446 \citep{Linares2012}, and $\eta > 1.35$ for 1RXS J180408.9$-$342058 \citep{Dohi+24}.}
Theoretical studies using various X-ray burst models (e.g., \cite{Lampe2016, Dohi+24}) suggest that $\eta$ depends on the physical properties of binary systems, although this remains under-explored. To further constrain the $\Delta t_{\rm rec}$--$F_{\rm per}$ relation, long-term monitoring of X-ray bursts is essential.

Due to their flexibility, CubeSats operating within the nanosatellite framework are well-suited for such long-term monitoring of X-ray bursts. NinjaSat, a 6U-size ($10 \times 20 \times 30~{\rm cm}^3$) CubeSat X-ray observatory, was launched into a sun-synchronous orbit at an altitude of approximately 530~km on 2023 November 11 
\citep{Enoto2020, Tamagawa2025PASJ}. 
Equipped with two Xe-based proportional counters (Gas Multiplier Counters; GMCs) covering the 2--50~keV energy range, NinjaSat's total effective area of approximately 32~cm$^2$ at 6~keV is over two orders of magnitude larger than the X-ray detectors on previously launched CubeSats \citep{Takeda2023JINST}.
This significant area allows for monitoring of persistent flux and detection of X-ray bursts, even within a CubeSat mission. NinjaSat enables agile follow-up observations of bright X-ray transients and facilitates extended long-term monitoring.

X-ray bursts from the new LMXB SRGA J144459.2$-$604207 (hereafter, SRGA J1444) are suitable targets for NinjaSat. 
SRGA J1444 was discovered in outburst by the Mikhail Pavlinsky ART-XC telescope onboard the Spectrum Roentgen Gamma (SRG) observatory on 2024 February 21 \citep{Molkov2024}, and later confirmed by MAXI \citep{2024ATel16469....1M} and Swift \citep{2024ATel16471....1C}. 
Prompt follow-up observations by NICER discovered the pulsation at 447.9~Hz and X-ray bursts, along with an orbital period at 5.22~h, identifying SRGA J1444 as an accretion-powered millisecond X-ray pulsar \citep{Ng2024}.
X-ray bursts from SRGA J1444 were also reported by Swift \citep{2024ATel16475....1M}, MAXI \citep{2024ATel16483....1N}, INTEGRAL \citep{2024ATel16485....1S}, Insight-HXMT \citep{2024ATel16548....1L}, and IXPE \citep{Papitto2024}. 
A quasi-periodic burst occurrence ranging from $\sim 1.7~{\rm hr}$ to $\sim 2.9~{\rm hr}$ was initially recognized in the INTEGRAL observations \citep{2024ATel16507....1S}, thus identifying SRGA J1444 as the sixth clocked burster. 
Subsequent IXPE observations reported the increase of recurrence time up to $\sim 7.9~{\rm hr}$ and the decrease of the persistent X-ray intensity, resulting in a significant low power-law index in equation~(\ref{eq:eta}), with $\eta \sim 0.8$ \citep{Papitto2024}. 
NinjaSat also observed SRGA J1444 and detected 12 X-ray bursts---the first detection of X-ray bursts by a CubeSat \citep{2024ATel16495....1T}.

In this letter, we report on the long-term monitoring campaign of SRGA J1444 conducted with the newly launched CubeSat X-ray observatory NinjaSat. Throughout this letter, all errors are given at the 1$\sigma$ confidence level unless otherwise specified.

\section{Observations and data reduction}







The GMC is a non-imaging gas X-ray detector equipped with a space-proven gas electron multiplier \citep{Tamagawa2009}.
Each GMC fits into a compact 1U-size (10 $\times$ 10 $\times$ 10~cm$^3$) with a mass of 1.2~kg, making it suitable for CubeSats.
The time-tagging resolution for each X-ray photon is 61~$\mu$s.
NinjaSat is capable of conducting the pointing observation of X-ray sources with an accuracy of less than $0\fdg1$, utilizing an X-ray collimator with a $2\fdg1$ field of view (full-width at half-maximum) equipped with each GMC.

Because of the high charged-particle background, we limit the GMC operation to a low-background region, with latitudes between $\sim 30^\circ$ south and $\sim 40^\circ$ north, excluding the South Atlantic Anomaly.
Currently, the astronomical operational area covers $\sim$~37\% of the total, although the observation efficiency for each source is also affected by the Earth occultation and the battery charging of the satellite.
In the payload commissioning phase, we observed the Crab Nebula for the detector calibration and successfully detected the pulsation at 33.8~ms, confirming that the absolute time is correctly assigned to each X-ray photon with an accuracy of at least sub-milliseconds \citep{Tamagawa2025PASJ}. 


NinjaSat observed SRGA J1444 for one day on 2024 February 23 (MJD 60363) and started a monitoring campaign on February 26  (MJD 60366), which continued through 2024 March 18 (MJD 60387) until the end of the outburst. 
The observation was conducted with one GMC (GMC1), whose detector calibration was more advanced at the end of the initial observation phase.
We also observed the Crab Nebula before and after the monitoring campaign of SRGA J1444, from 2024 February 23 to February 26 and on March 19.
The detector calibration with the Crab Nebula indicates that using background data from the same observation period is more appropriate than using blank sky data from different periods under the current background model.
Therefore, we estimated the background level using data from the SRGA J1444 observation period when the satellite was not pointed at either SRGA J1444 or the Earth.
We got effective exposures of 197.5~ks, 104.7~ks, and 16.6~ks for SRGA J1444, background data, and the Crab Nebula, respectively. 
Photon arrival times were corrected to the solar system barycenter with FTOOLS {\tt barycen} using the DE-405 planetary ephemeris with source coordinates R.A. $= 221\fdg24558$, Decl.$= - 60\fdg69869$ obtained by Chandra \citep{2024ATel16510....1I}.


The main background component in the GMC data is the non-X-ray background, which includes charged particle events and electrical noise events.
The GMC has two readout pads: a circular inner readout electrode with a radius of 25.0 mm and an annular outer electrode with radii between 25.1 mm and 33.5 mm.
Signals from each readout pad are digitized by a 12-bit analog-to-digital converter at a sampling rate of 25~MHz, followed by onboard analysis to extract waveform parameters---such as pulse height, rise time, and the Pearson correlation coefficient between the inner and outer waveforms $R$---which are subsequently downlinked.
Charged particle events entering perpendicular to the readout pad can be rejected because their signal rise time ($\sim$ 1~$\mu$s) is longer than that of X-ray events ($\sim 400$~ns) due to the difference in the distributed length of the electron cloud along the drift direction.
On the other hand, events from the parallel direction leave signals on both pads, resulting in a relatively high correlation coefficient $R$. 
This allows them to be distinguished from X-ray events, except when X-rays enter between the pads.
Additionally, the event cut based on the correlation coefficient $R$ is also useful for the common-mode electrical noise simultaneously triggered in both channels, which typically has a value of $R \sim 0.95$.

In this letter, we only use the X-ray count rate for the following analysis without the response for the spectral discussions.
The persistent count rate in the 2--10~keV band is converted to X-ray intensity in units of mCrab, i.e., a flux referenced to the count rate of the Crab Nebula.
Because the detector calibration is still ongoing, we employed a tentative event selection criterion: selecting events with signal rise times in the range of 200–800~ns and correlation coefficients $R$ of less than 0.4. 
The average raw background rates of the inner and outer regions in the 2--10~keV band are 6.0~counts~s$^{-1}$ and 10.6~counts~s$^{-1}$, respectively.
After applying the event cut, the rates are reduced to 0.294 $\pm$ 0.003~counts~s$^{-1}$ and 2.26 $\pm$ 0.02~counts~s$^{-1}$, respectively.
\textcolor{black}{
The event cut based on the correlation coefficient $R$ between the inner and outer waveforms is sensitive to events that are simultaneously triggered in both regions. Consequently, it functions similarly to the anti-coincidence method, making it more effective in the inner region, which is surrounded by the outer region.
}
Thus, we analyzed only event data of the inner region for the persistent flux evaluation to obtain a higher signal-to-noise ratio, while we used event data extracted from both regions for the X-ray burst analysis.
The background-subtracted rate of the Crab Nebula in the 2--10~keV band is estimated to be 11.89 $\pm$ 0.03~counts~s$^{-1}$.
The time variation in the 3.0-hr binned background light curve shows a slight periodicity at approximately 1 and 6 days and follows a Gaussian distribution with a 1$\sigma$ width of 0.025~counts~s$^{-1}$, corresponding to 2.5~mCrab.

\section{Data analysis and results} 

To search for X-ray bursts, we extracted 10-s binned light curves from all screening data and clearly detected 12 X-ray bursts (IDs 1--12), as listed in table~\ref{tab:burst_prop}.
These bursts exhibit a peak X-ray intensity of $\sim 1$~Crab and lasting $\sim$ 20~s \citep{2024ATel16495....1T}, consistent with observations reported by NICER and SRG \citep{Ng2024, Molkov2024}.

\begin{figure}
    \begin{center}
    \includegraphics[width=85mm]{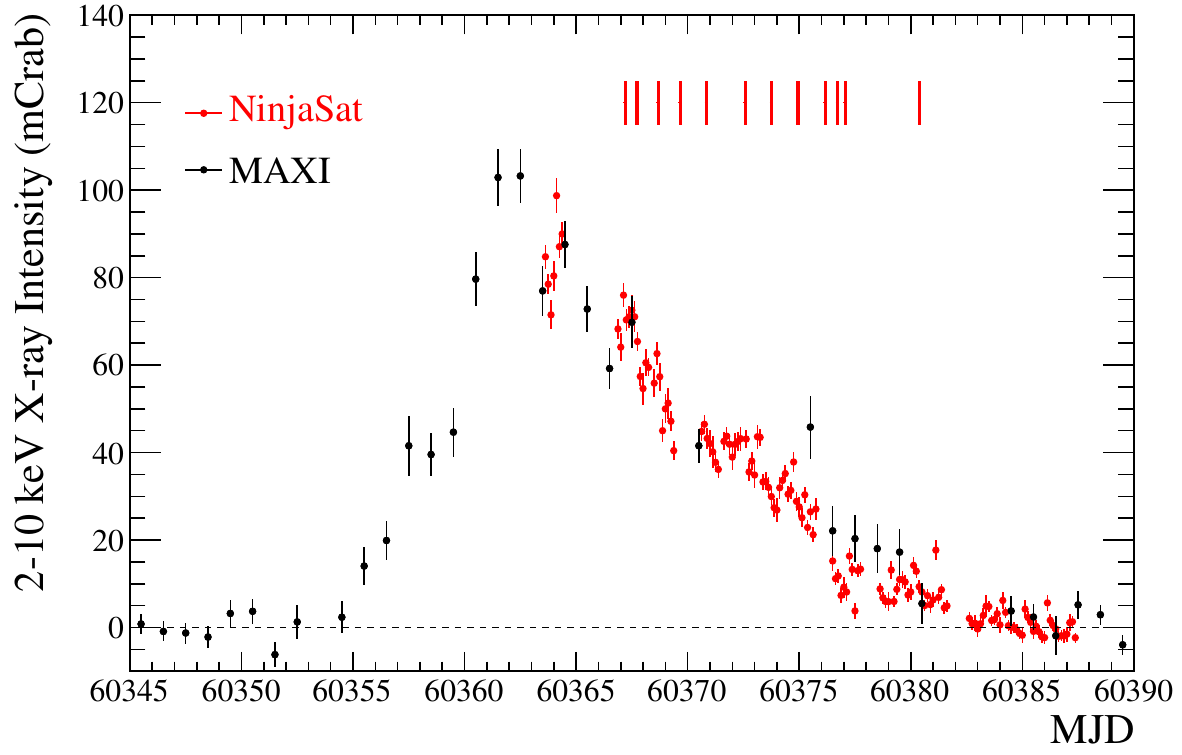}
    \end{center}
    \caption{2--10~keV light curves of SRGA J1444 monitored by NinjaSat (red) and MAXI (black) with the binsizes of 3.0 hr and 24 hr, respectively. 
    \textcolor{black}{The NinjaSat light curve is calculated using event data from the inner region with a subtraction of the averaged background rate (0.294~counts~s$^{-1}$).}
    The time intervals between 5-s before and 50-s after the burst onset are excluded to show the persistent flux decay, while these burst onsets are indicated by red vertical lines.
    {Alt text: Time series plot.} 
    }
    \label{fig:SRGA_lightCurve}
\end{figure}

Figure \ref{fig:SRGA_lightCurve} shows the persistent light curve of SRGA J1444 in the 2–10~keV band observed with NinjaSat, compared with that from MAXI \citep{Matsuoka2009}.
The 2--10~keV X-ray intensity observed with MAXI was calculated using publicly available data.\footnote{http://maxi.riken.jp/top/slist.html}
The outburst of SRGA J1444 reached the maximum X-ray intensity at $\sim$ 100~mCrab at MJD 60361 and then gradually decayed for nearly 30 days to the background level. 
The NinjaSat monitoring campaign covered almost the entire outburst decay phase until MJD 60387.
The flux evolution observed with NinjaSat is consistent with that of MAXI, achieving comparable statistical errors with 3.0-hr bins to those of the MAXI daily light curve.
NinjaSat enables us to track finer variations in the persistent X-ray intensity.

{
\tabcolsep = 5.1pt

\begin{table*}
\tbl{Properties of Type-I X-ray bursts from SRGA J1444 observed with NinjaSat.}{
\scalebox{1}{
\begin{tabular}{l*{10}{c}}
\hline
ID & 
MJD \footnotemark[$*$] & 
\begin{tabular}{c} $\Delta t_{\rm pre}$\footnotemark[$\dagger$] \\(hr) \end{tabular} & 
\begin{tabular}{c} $t_{\rm  rise}$\footnotemark[$\S$] \\(s)\end{tabular} & 
\begin{tabular}{c} $t_{\rm pl}$\footnotemark[$\|$] \\(s)\end{tabular} & 
\begin{tabular}{c} $\tau_{\rm D}$\footnotemark[$\#$] \\(s) \end{tabular}& 
\begin{tabular}{c} $A$\footnotemark[$**$] \\(counts~s$^{-1}$) \end{tabular}& 
\begin{tabular}{c} Fluence\\(counts) \end{tabular}& 
\begin{tabular}{c} $\tau$\footnotemark[$\dagger\dagger$] \\(s) \end{tabular}& 
\begin{tabular}{c} $\Delta t_{\rm rec}$\footnotemark[$\S\S$] \\(hr) \end{tabular}& 
\begin{tabular}{c} $\chi^2$/d.o.f. \end{tabular}\\
\hline
1 & 60367.19877 & - & 5.3 $\pm$ 0.6 & 9.6 $\pm$ 0.8 & 5.8 $\pm$ 1.5 & 12.0 $\pm$ 1.1 & 216 $\pm$ 29 & 18.1 $\pm$ 2.9 & - & 82.1/105 \\
2 & 60367.73134 & 12.782 & 0.6 $\pm$ 1.7 & 10.2 $\pm$ 2.1 & 11.9 $\pm$ 2.0 & 12.9 $\pm$ 1.3 & 289 $\pm$ 48 & 22.4 $\pm$ 4.3 & 2.130 & 107.8/103 \\
3 & 60368.68673 & 22.929 & 3.4 $\pm$ 1.4 & 13.6 $\pm$ 1.4 & 3.0 $\pm$ 1.0 & 11.4 $\pm$ 1.0 & 209 $\pm$ 29 & 18.3 $\pm$ 3.0 & 2.293 & 82.1/102 \\
4 & 60369.67668 & 23.759 & 5.2 $\pm$ 0.7 & 4.0 $\pm$ 1.1 & 8.8 $\pm$ 1.5 & 14.1 $\pm$ 1.5 & 216 $\pm$ 35 & 15.4 $\pm$ 3.0 & 2.970 & 103.4/102 \\
5 & 60370.85748 & 28.339 & 1.4 $\pm$ 1.6 & 3.5 $\pm$ 1.9 & 9.4 $\pm$ 1.4 & 18.2 $\pm$ 2.3 & 248 $\pm$ 54 & 13.6 $\pm$ 3.4 & 3.149 & 113.7/104 \\
6 & 60372.57310 & 41.175 & 3.5 $\pm$ 0.8 & 10.4 $\pm$ 1.3 & 6.0 $\pm$ 1.1 & 13.3 $\pm$ 1.2 & 241 $\pm$ 31 & 18.2 $\pm$ 2.9 & 3.167 & 91.2/102 \\
7 & 60373.76358 & 28.571 & 1.9 $\pm$ 0.4 & 9.4 $\pm$ 1.0 & 5.7 $\pm$ 1.4 & 14.7 $\pm$ 1.3 & 237 $\pm$ 33 & 16.0 $\pm$ 2.6 & 3.571 & 101.0/103 \\
8 & 60374.94451 & 28.342 & 0.3 $\pm$ 1.3 & 10.4 $\pm$ 1.7 & 6.2 $\pm$ 1.3 & 14.1 $\pm$ 1.3 & 235 $\pm$ 39 & 16.7 $\pm$ 3.2 & 4.049 & 116.4/103 \\
9 & 60376.16844 & 29.374 & 0.5 $\pm$ 0.1 & 8.0 $\pm$ 1.3 & 9.3 $\pm$ 2.3 & 13.6 $\pm$ 1.3 & 239 $\pm$ 43 & 17.6 $\pm$ 3.6 & 4.896 & 114.7/102 \\
10 & 60376.72939 & 13.463 & 0.5 $\pm$ 0.8 & 9.6 $\pm$ 1.3 & 5.2 $\pm$ 1.2 & 16.1 $\pm$ 1.4 & 244 $\pm$ 36 & 15.1 $\pm$ 2.6 & 6.731 & 110.5/100 \\
11 & 60377.05892 & 7.909 & 0.5 $\pm$ 1.0 & 7.2 $\pm$ 1.2 & 5.1 $\pm$ 1.2 & 18.0 $\pm$ 1.7 & 227 $\pm$ 39 & 12.6 $\pm$ 2.4 & 7.909 & 111.2/103 \\
12\footnotemark[$\|\|$] & 60380.40925 & 80.408 & - & - & - & - & - & - & 10.051 & - \\
\hline
1--11 & - & - & 1.4 $\pm$ 0.4 & 9.8 $\pm$ 0.6 & 7.7 $\pm$ 0.5 & 13.3 $\pm$ 0.4 & 241 $\pm$ 13 & 18.1 $\pm$ 1.2 & - & 103.9/105 \\
1--3 & - & - & 4.4 $\pm$ 0.7 & 10.9 $\pm$ 1.1 & 5.3 $\pm$ 0.9 & 12.0 $\pm$ 0.7 & 232 $\pm$ 23 & 19.3 $\pm$ 2.3 & - & 91.4/105 \\
4--6 & - & - & 4.2 $\pm$ 0.6 & 4.9 $\pm$ 1.4  & 8.9 $\pm$ 0.9 & 15.1 $\pm$ 1.2 & 241 $\pm$ 32 & 15.9 $\pm$ 2.4 & - & 122.5/105 \\
7--8 & - & - & 2.5 $\pm$ 0.4 & 9.0 $\pm$ 0.7 & 5.9 $\pm$ 0.7 & 14.3 $\pm$ 0.8 & 230 $\pm$ 19 & 16.1 $\pm$ 1.6 & - & 106.1/105 \\
9--11 & - & - & 0.3 $\pm$ 0.3 & 5.8 $\pm$ 1.3 & 9.6 $\pm$ 1.2 & 17.0 $\pm$ 1.2 & 264 $\pm$ 35 & 15.5 $\pm$ 2.3 & - & 145.1/105 \\
\hline
\end{tabular}}
}
\label{tab:burst_prop}
\begin{tabnote}
\vspace*{-0.25cm}
\begin{multicols}{2}
    \footnotemark[$*$] MJD: burst onset time in Modified Julian Date (MJD).\\
    \footnotemark[$\dagger$] $\Delta t_{\rm pre}$: elapsed time (hr) since the previous burst detected with NinjaSat. \\
    \footnotemark[$\S$] $t_{\rm rise}$: time to reach the peak from the onset in a unit of seconds. \\
    \footnotemark[$\|$] $t_{\rm pl}$: duration of the plateau (s). \\
    \footnotemark[$\#$] $\tau_{\rm D}$: decay time constant (s). \\
    
    \footnotemark[$**$] $A$: burst amplitude, i.e., the count rate during the plateau (counts s$^{-1}$). \\
    \footnotemark[$\dagger\dagger$] $\tau$: equivalent duration (s), ratio of burst integrated fluence to peak flux. \\
    \footnotemark[$\S\S$] $\Delta t_{\rm rec}$: average burst recurrence time (hr) (see section~\ref{sec:MCMC}). \\
    \footnotemark[$\|\|$] Only the burst onset times, $\Delta t_{\rm pre}$, and $\Delta t_{\rm ave}$ are listed because the burst was truncated by the boundary of the observations.\\
    \end{multicols}

    \vspace*{-0.35cm}
\end{tabnote}
\end{table*}
}

\subsection{Evolution of burst profiles} 
\label{sec:profile}

\begin{figure}[t]
    \begin{center}
        \includegraphics[width=90mm]{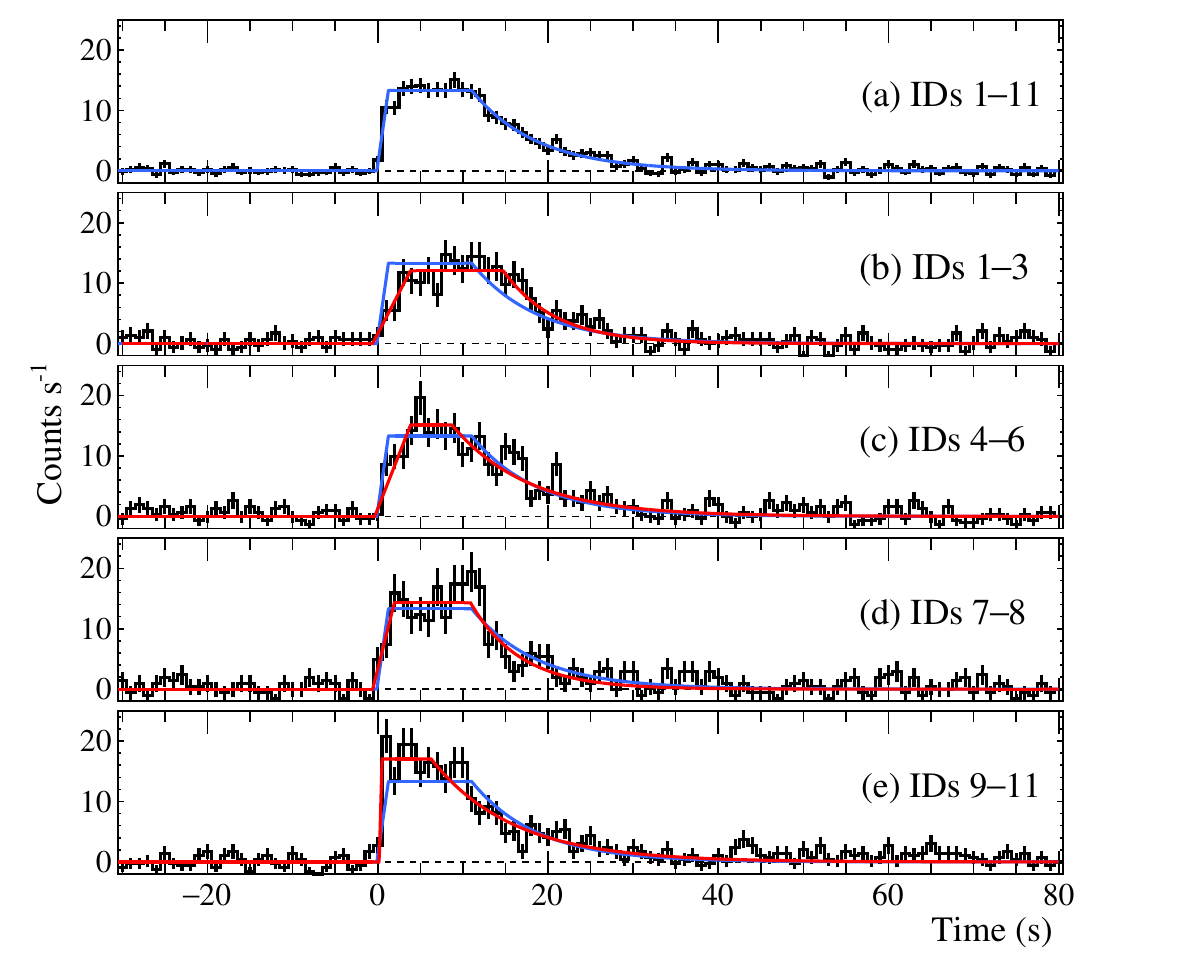}
    \end{center}
    \caption{Average profiles of X-ray burst IDs 1--11, 1--3, 4--6, 7--8, and 9--11 (top to bottom). These light curves are calculated in the 2--20 keV energy band at 1 s resolution after subtraction of the persistent emission based on fitting results. The blue solid line represents the best-fit model of the linear rise, plateau, and exponential decay (see section \ref{sec:profile}) applied to the overall average profile~(a). In the lower four panels, the red solid lines are the best-fit models for each burst profile~(b)--(e), where the best-fit average profile (blue solid line) is also shown for comparison.
    The best-fit parameters are summarized in table \ref{tab:burst_prop}.
    {Alt text: Five line graphs.} 
    }
    \label{fig:SRGA_burstLC_average}
\end{figure}

\begin{figure}
    \begin{center}
        \includegraphics[width=85mm]{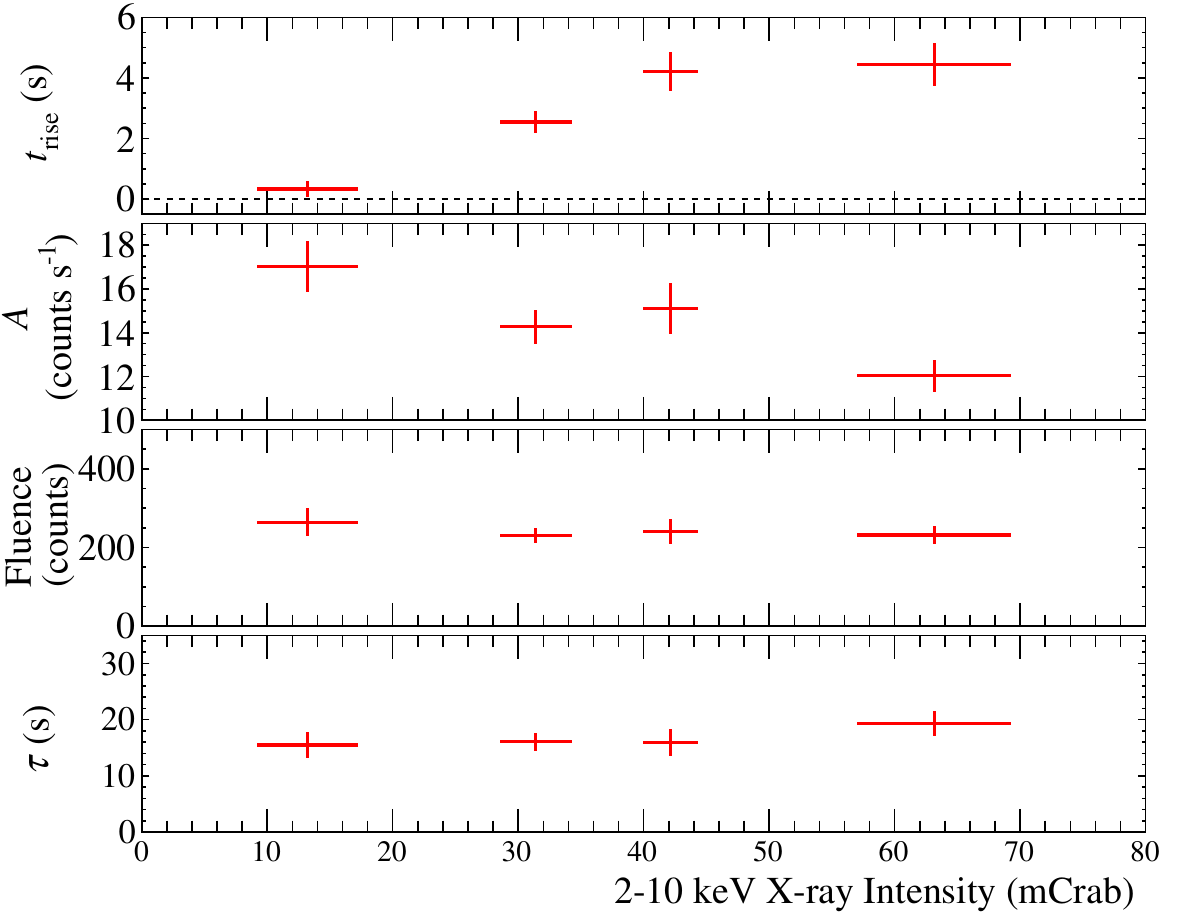}
    \end{center}
    \caption{Flux dependence of the various parameters in burst profiles: time to reach the peak flux $t_{\rm rise}$, burst amplitude $A$, burst fluence, and equivalent duration $\tau$ (top to bottom). The X-axis represents the persistent X-ray intensity estimated by linear interpolation and averaging the NinjaSat light curves at each interval (IDs 1–3, 4–6, 7–8, and 9–11).
    {Alt text: Four scatter plots.} 
    }
    \label{fig:burstParamters_vs_flux}
\end{figure}

Figure~\ref{fig:SRGA_burstLC_average}(a) shows the overall averaged burst profile in the 2--20~keV band.
The burst profile is characterized by a fast linear rise, followed by a plateau and an exponential decay, similar to those reported in NICER and SRG observations \citep{Ng2024, Molkov2024}.
To investigate the evolution of the burst profile during the outburst with better statistics, we combined the first 11 bursts into four intervals based on the persistent X-ray intensity. Then we fitted each light curve with a burst model ($f(t)$ as a function of time, $t$), described by
{
\fontsize{8}{13}
\begin{equation}
    \begin{array}{l}
    \leftline{$f(t)$ =} \\
        \begin{cases}
            c_{\rm per} & t \leq t_0 \\
            \frac{A}{t_{\rm rise}}\left(t-t_0\right) + c_{\rm per} & t_0 < t \leq t_0 + t_{\rm rise} \\
            A + c_{\rm per} & t_0 + t_{\rm rise} < t \leq t_0 + t_{\rm rise} + t_{\rm pl} \\
            A \exp\left(-\frac{t-t_0 - t_{\rm rise} - t_{\rm pl}}{\tau_{D}}\right) + c_{\rm per} & t > t_0 + t_{\rm rise} + t_{\rm pl}
        \end{cases},
    \end{array}
    \label{eq:profile_burst}
\end{equation}
}where $c_{\rm per}$ is the persistent rate ($\rm counts~s^{-1}$), $A$ is the burst amplitude ($\rm counts~s^{-1}$), $t_0$ is the burst onset time (s), $t_{\rm rise}$ is the time to reach the peak from the onset (s), $t_{\rm pl}$ is the duration of the plateau (s), and $\tau_{D}$ is the decay time constant (s). 
Figures \ref{fig:SRGA_burstLC_average}(b)--(e) show the averaged profiles at each interval (IDs 1--3, 4--6, 7--8, and 9--11) with the best-fit models.
The corresponding best-fit parameters are given in table \ref{tab:burst_prop}. 
The X-ray burst profiles showed a significant evolution in morphology; The burst rise time ($t_{\rm rise}$) became faster, and its amplitude ($A$) increased as the outburst decayed.
In addition, we evaluated the burst fluence using the fit results and then employed the equivalent burst duration $\tau$, which is defined as the ratio of burst fluence to peak intensity, as a useful indicator independent of the uncertainty of the distance.
Figure \ref{fig:burstParamters_vs_flux} shows the dependence of $t_{\rm rise}$, $A$, the burst fluence, and $\tau$ on the persistent level, which are estimated by linear interpolation and averaging the NinjaSat light curve (figure \ref{fig:SRGA_lightCurve}) at each interval. 
While the persistent level, which is proportional to the mass accretion rate, decreased from approximately 63~mCrab to 13~mCrab, the rise time decreased from $t_{\rm rise}=4.4\pm0.7$~s to $t_{\rm rise} = 0.3\pm0.3$~s, and the amplitude increased by 44\% from $ A = 11.8\pm0.7$~$\rm counts~s^{-1}$ to $A = 17.0\pm1.2$~$\rm counts~s^{-1}$. 
In contrast, the fluence showed no significant changes, with average values of 240~counts.
The equivalent duration marginally decreased from $\tau = 19.3 \pm 2.3$~s to $\tau = 15.5 \pm 2.3$~s.
The best-fit parameters for each burst are also listed in table~\ref{tab:burst_prop}.
The burst onset times are determined with an accuracy of approximately 1~s, and the elapsed times since the previous burst $\Delta t_{\rm pre}$ range from $7.909$~hr to $80.408$~hr. 



\textcolor{black}{
We assessed the systematic uncertainties in our burst profile analysis based on count rates without performing spectral analysis.
\citet{Fu2024arXiv241205779F} reported Insight-HXMT's observations of photospheric radius expansions (PREs) in 14 out of a total of 60 bursts.
They presented the evolution of blackbody temperature, which is kept high (roughly 2--3 keV) in spite of the occurrence of PREs. 
This fact justifies our approach using the count rate rather than the bolometric flux.
This stems from a small variation of count rate to bolometric flux conversion factor by only $\pm$15\% during bursts, given the effective area of the GMC, which is comparable to the 1$\sigma$ statistical error of light curves as shown in figure~\ref{fig:SRGA_burstLC_average}. 
Furthermore, since the PRE effects on the spectrum are remarkable in a short timescale of less than $\sim$3~s, they little affect characteristic parameters such as $\tau$ in an overall burst.
}

\subsection{MCMC inference of burst recurrence time as a function of persistent flux}
\label{sec:MCMC}

During the campaign, NinjaSat monitored the evolution of persistent X-ray intensity and detected 12 X-ray bursts (figure~\ref{fig:SRGA_lightCurve}).
Based on a comparison with IXPE observations \citep{Papitto2024}, only burst IDs 10 and 11 were confirmed to be consecutive, with the burst recurrence time of $\Delta t_{\rm rec} = 7.9$~hr.
To quantify the $\Delta t_{\rm rec}$--$F_{\rm per}$ relation in SRGA J1444, we developed a new method using a Markov chain Monte Carlo (MCMC) approach, which is applicable even when several bursts fall within observation gaps and are consequently missed.



\subsubsection{MCMC method} 
The empirical $\Delta t_{\rm rec}$--$F_{\rm per}$ relation in equation~(\ref{eq:eta}) indicates the existence of the conserved quantity $C$ for a pair of bursts.
When the persistent flux varies substantially between bursts, equation~(\ref{eq:eta}) cannot be applied directly to the observed data and must instead be transformed into an integral form. 
Moreover, due to incomplete observational coverage, bursts are often missed, as in the case of the NinjaSat observations.
Even in such cases, the integral of $F_{\rm per}^{\eta}$ over the interval between burst detection times should be equal to $C$ multiplied by \textcolor{black}{the actual number of burst-to-burst intervals, $n_i$, between two detected bursts}:
\begin{equation}
    \int_{t_i}^{t_{i+1}} F_{\rm per}^{\eta} {\rm d}t = n_iC,
    \label{eq:eta_miss}
\end{equation}
where $t_i$ and $t_{i+1}$ are the burst onset times.
We employed an MCMC algorithm to estimate the parameters in equation (\ref{eq:eta_miss}) that best match the observations.
Bayesian statistics using MCMC methods have recently been applied to a wide range of fields in astrophysics, including modeling X-ray bursts demonstrated in several studies (e.g., \cite{Goodwin2019, Johnston2020, Galloway2024}). We used a discretized form of equation~(\ref{eq:eta_miss}) to evaluate the simple Gaussian likelihood function, $p(D|\eta, C, n_i)$, by comparing predicted values with the observational data, $D$, using 
\begin{equation}
    p(D|\eta, C, n_i) =\\ \prod_{i}\frac{1}{\sqrt{2\pi \sigma_i^2}} \exp \left[-\frac{(\sum F_{\rm per}^{\eta} t_{\rm bin} - n_iC)^2}{2\sigma_i^2}\right],
\end{equation}
where $F_{\rm per}$ is measured in units of the overall averaged X-ray intensity of $26$~mCrab, $t_{\rm bin}$ is the time width to interpolate the persistent flux in units of 1 day, $\sigma_i$ is the observational error, and $i\ (=1,2,\ldots,11)$ is iterated over each observed burst.

For the MCMC calculations, we used the open-source Python package {\tt emcee} \citep{Foreman2013}.
To determine the persistent fluxes outside observation intervals, we linearly interpolated the values between observations with a time bin of $t_{\rm bin} =$ 1/864~\textcolor{black}{day (= 100~s)}.
We employed flat prior distributions for $\eta$ and $C$, setting broad acceptable ranges\footnote{The range of $C$ was set based on its maximum and minimum values obtained from two consecutive bursts (IDs 10--11), where assuming $\eta$ range of 0.5--1.5.}: $0.5 \leq \eta \leq 1.5$ and $0.1 \leq C \leq 0.25$.
Although $n_i$ is expected to be close to integer values, slight variations in $C$ from burst to burst shift the posterior distributions of $n_i$ away from integers.
Therefore, for the prior of $n_i$, we employed a model based on multiple Gaussian functions, each centered on an integer value in the range of 1--20 with a standard deviation $\sigma$ set so that $3\sigma = 0.5$.
As an exceptional case, the center value was \textcolor{black}{set} to 1 for $n_{10}$ because burst IDs 10 and 11 are consecutive, as described above. 
Additionally, we impose another constraint on the average burst recurrence times $\Delta t_{\rm rec} \equiv \Delta t_{\rm pre}/n_i$, setting $\Delta t_{\rm rec}> 2$~hr. This is based on the report that $\Delta t_{\rm rec}$ after MJD 60367, corresponding to the NinjaSat burst detection period, was longer than 2~hr \citep{Molkov2024}.

We ran the MCMC chains with 200 walkers for $2\times10^5$ steps. 
The walkers were uniformly initialized within the ranges of the flat prior distributions to comprehensively explore the $(\eta, C, n_i)$-parameter space.
Given the possibility that the sampled distribution could be multimodal, we used a combination of moves, DEMove and DESnookerMove, with weights of 80\% and 20\%, respectively, as suggested in the {\tt emcee} documentation.\footnote{https://emcee.readthedocs.io/en/stable/}

\subsubsection{Results}

The integrated autocorrelation time $\tau$ is a reliable indicator for assessing the convergence of the MCMC chain.
In {\tt emcee}, running the chain for 50$\tau$ samples generally ensures the convergence\footnotemark[5].
We estimated $\tau$ using steps from the last half of the total and found $\tau \sim 1300$.
Therefore, the initial $1.8\times10^5$ steps in each chain were discarded as burn-in to ensure full convergence.

\begin{figure}
    \begin{center}
        \includegraphics[width=85mm]{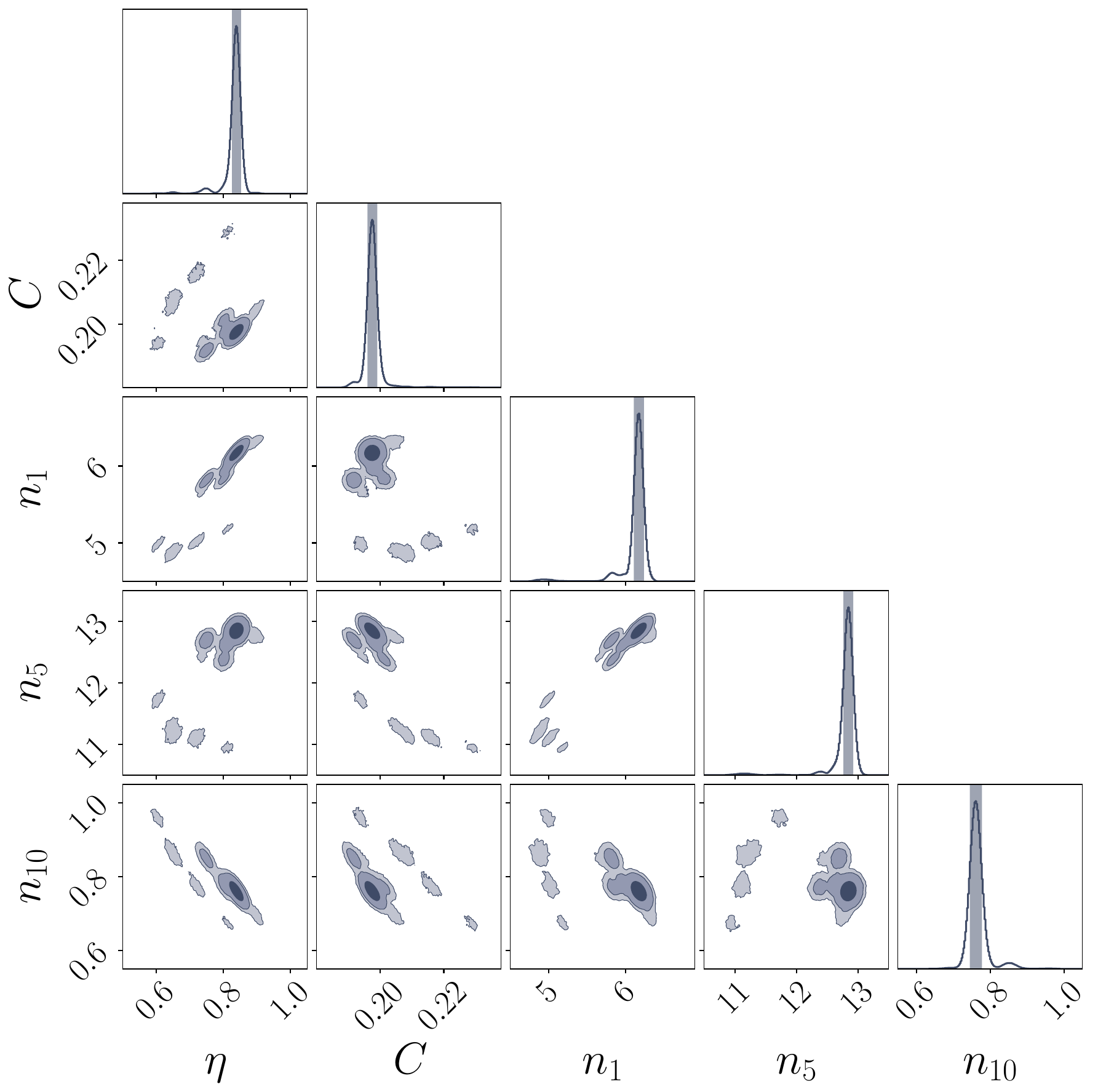}
    \end{center}
    \caption{Two-dimensional posterior distributions of power-law index $\eta$, $C$, and $n_i\ (i = 1, 5, 10)$ with marginalized histograms along the diagonal. The confidence contours displayed in each 2-D panel correspond to 1, 2, and 3$\sigma$ levels. The vertical bands in the marginalized histograms along the diagonal are the inferred 68\% highest posterior density intervals.
    {Alt text: Ten contour plots and five histograms.} 
    }
    \label{fig:MCMC_posterior}
\end{figure}

Two-dimensional posterior distributions of $\eta$, $C$, and $n_i\ (i = 1, 5, 10)$ are shown in figure~\ref{fig:MCMC_posterior}, with marginalized histograms along the diagonal. 
Each parameter space exhibits multi-modal distributions, which can be attributed to the fact that the only strictly constrained value is $n_{10} \sim 1$. 
This allows for multiple combinations of $\eta$ and $C$ that result in $n_i$ values close to integers.
Nevertheless, it is also evident that each parameter has a prominent peak.
To evaluate the prominent peaks in multi-modal distributions, we calculate the highest posterior density (HPD) intervals, which are particularly suited for cases with multimodality or asymmetry (e.g., \cite{Gelman2014}), using the {\tt hdi} function from {\tt ArviZ}---a Python package for exploratory analysis of Bayesian models \citep{Kumar2019}---with the 'multimodal' option.
The maximum likelihood estimates for each 1-D marginalized posterior, with 68\% HPD intervals, are listed in table~\ref{tab:MCMC_estimate}.
The inferred power-law index is significantly lower than 1, $\eta = 0.84^{+0.02}_{-0.01}$.
The ratios of the maximum likelihood estimates $n_i$ to their nearest integer values are well within $\pm$5\%, except for $n_{10}$, which exhibits a residual of approximately 25\%.
The average burst recurrence times for each observed burst $\Delta t_{\rm rec}$ increased from $2.1$~hr to $10.1$~hr, as given in table~\ref{tab:burst_prop}. 

{
\tabcolsep = 2.0pt

\begin{table*}
  \tbl{Maximum likelihood estimates for each 1D marginalized posterior distribution with 68\% highest posterior density intervals.}{%
  \fontsize{6.8}{11}\selectfont
  \begin{tabular}{*{13}{c}}
  \hline
      $\eta$ & $C$ & $n_1$ & $n_2$ & $n_3$ & $n_4$ & $n_5$ & $n_6$ & $n_7$ & $n_8$ & $n_9$ & $n_{10}$ & $n_{11}$ \\
  \hline
  \fontsize{6.8}{20}\selectfont
    $0.84^{+0.02}_{-0.01}$ & 
    $0.198^{+0.002}_{-0.002}$ & 
    $6.16^{+0.07}_{-0.06}$ & 
    $9.70^{+0.10}_{-0.08}$ & 
    $8.15^{+0.06}_{-0.05}$ & 
    $9.05^{+0.06}_{-0.06}$ & 
    $12.84^{+0.09}_{-0.07}$ & 
    $8.16^{+0.06}_{-0.05}$ & 
    $7.07^{+0.05}_{-0.05}$ & 
    $6.10^{+0.05}_{-0.05}$ & 
    $1.93^{+0.02}_{-0.03}$ & 
    $0.76^{+0.02}_{-0.02}$ & 
    $7.96^{+0.14}_{-0.13}$ \\
  \hline
  \end{tabular}}
  \label{tab:MCMC_estimate}
  \begin{tabnote}
  \end{tabnote}
\end{table*}
}


\
\vspace{-15pt}

\subsection{Recurrence time variation with burst duration}
\begin{figure}[t]
    \begin{center}
        \includegraphics[width=85mm]{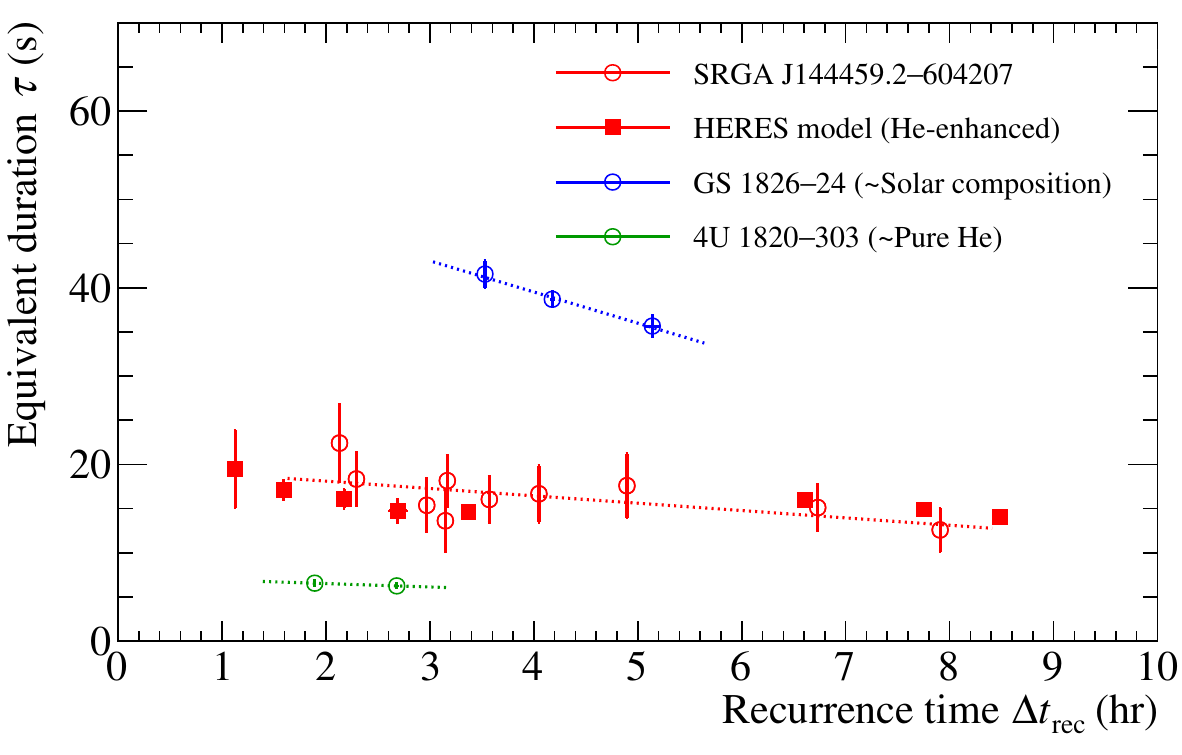}
    \end{center}
    \caption{Relation between the burst equivalent duration $\tau$ and the recurrence time $\Delta t_{\rm rec}$, with the best-fit linear model shown as a red dotted line. 
    Eight filled-circles are calculated from our model HERES in case of $X/Y=1.5$, $Z_{\rm CNO} = 0.015$, and the accretion rate of $\dot{M} =\left(0.8\text{--}5\right)\times 10^{-9}~M_{\odot}~{\rm yr}^{-1}$ \citep{Dohi2025PASJL}, where $M_{\odot}$ denotes the solar mass.
    The observed relations for GS 1826$-$24 and 4U 1820$-$303 are also shown in blue and green, respectively, taken from \citet{Galloway2017}. The best-fit parameters are as follows: SRGA J1444 $(a=-0.8 \pm 0.5{\rm~s~hr^{-1}},\ b = 20 \pm 2~{\rm s},\ \chi^2/{\rm d.o.f}=3.22/8)$, GS 1826$-$24 
    $(a = -3.5 \pm 1.1{\rm~s~hr^{-1}},\ b = 54 \pm 5~{\rm s},\ \chi^2/{\rm d.o.f} = 0.13/1)$
    , and 4U 1820$-$303 $(a = -0.4{\rm~s~hr^{-1}},\ b = 7.3~{\rm s})$.
    {Alt text: Graph showing the scatter plots with best-fit linear models.} 
    }
    \label{fig:tau_vs_deltaT}
\end{figure}

The compositions of the accreted matter, i.e., mass fractions of hydrogen ($X$), helium ($Y$), and heavier CNO elements or metallicity ($Z_{\rm CNO}$), are reflected in the burst properties such as the burst equivalent duration $\tau$ and the recurrence time $\Delta t_{\rm rec}$~\citep{Lampe2016}.
Although only an upper limit of 10.6~kpc on the source distance has been determined for SRGA J1444 \citep{Ng2024}, both $\tau$ and $\Delta t_{\rm rec}$ are independent of the distance, making these parameters useful for comparing observations and theoretical predictions across different sources.
Figure~\ref{fig:tau_vs_deltaT} shows the $\tau$--$\Delta t_{\rm rec}$ relation of SRGA J1444, alongside the clocked burster GS 1826$-$24, which has a near-solar composition (e.g., \cite{Johnston2020}), and the ultra-compact binary 4U 1820$-$303, believed to be a pure-He burster with a low-mass He white dwarf~(e.g., \cite{Cumming2003, Galloway2008}). 
The burst recurrence times $\Delta t_{\rm rec}$ of GS 1826$-$24 and 4U 1820$-$303 are taken from table~2 in \citet{Galloway2017}, and $\tau \equiv E_b/F_{\rm pk}$ are calculated using the burst fluence $E_b$ and peak flux $F_{\rm pk}$ from the same table.
We also show the theoretical relation from our model HERES with $X/Y=1.5$ and $Z_{\rm CNO}=0.015$ (for the HERES model, see \cite{2020PTEP.2020c3E02D}), which is in line with various observations of SRGA J1444 \citep{Dohi2025PASJL}.
To characterize the observed $\tau$--$\Delta t_{\rm rec}$ relations, we fitted a linear model ($\tau = a\Delta t_{\rm rec} + b$) to each data.
The average burst duration of SRGA J1444 is less than half that of GS 1826$-$24 and three times that of 4U 1820$-$303.
Furthermore, the negative slope between $\tau$ and $\Delta t_{\rm rec}$ becomes shallower for sources with shorter $\tau$ values.

\vspace{-0.3cm}

 \section{Discussion and conclusion}

In this letter, we present the long-term monitoring of SRGA J1444 conducted with NinjaSat over a period of approximately 25 days.
We found that SRGA J1444 exhibited X-ray bursts with a fast rise time of $<$5~s and a short duration of \textcolor{black}{$\tau = 18.1 \pm 1.2$~s (IDs 1--11), the latter of which is consistent with the values derived by the spectral analysis from other satellites, such as IXPE ($\tau = 16.8 \pm 1.6$~s, \cite{Papitto2024}) and SRG ($\tau \sim 16$~s, \cite{Molkov2024}), with an accuracy of $\sim$10\%.}
The fast rise time and short duration are characteristic features of sources with relatively He-rich accreted fuel.
This is because He burns during the burst via the triple-$\alpha$ reaction, which proceeds on a much shorter time scale than the H burning via the hot-CNO cycle, $rp$, and $\alpha p$ processes.
The observed burst duration in SRGA J1444 is longer than in pure-He bursters but shorter than sources with the solar composition (figure~\ref{fig:tau_vs_deltaT}).
Given that no photospheric radius expansion burst has been observed from SRGA J1444, including with observations from other satellites, it is reasonable to assume that SRGA J1444 has a relatively He-enhanced accreted fuel compared to the solar composition. 
The He-enhanced scenario with $X/Y=1.5$ and $Z_{\rm CNO}=0.015$ is further supported by the theoretical predictions from the HERES model \citep{Dohi2025PASJL}.

We show that the recurrence time in SRGA J1444 is roughly inversely proportional to the persistent X-ray intensity with a power-law index $\eta = 0.84^{+0.02}_{-0.01}$. 
We also noted that $n_i$ are closely matched to integer values within $\pm$5\%, except for $n_{10}$.
The deviations from integers could be attributed to the average variation in $C$ between each detected burst and/or to limitations of the simplified burst model expressed in equation~(\ref{eq:eta_miss}). 
Contributing factors may include incomplete observational coverage, potential variations in persistent flux on timescales of 3.0 hr or shorter, which corresponds to the time resolution of the light curve (figure~\ref{fig:SRGA_lightCurve})\textcolor{black}{, and long-term spectral variations associated with the outburst.}
\textcolor{black}{Our results are consistent with the IXPE observation reported by \citet{Papitto2024}, which shows $\eta \sim 0.8$ with recurrence time deviations ranging from a few \% to roughly 10\%\footnote{The variation of the recurrence time is reflected in the deviation of $n_i$ from their nearest integer values in our formulation expressed in equation~(\ref{eq:eta_miss}).} (see also Fu et al. 2024).
}


The estimated index $\eta = 0.84^{+0.02}_{-0.01}$ of SRGA J1444 is the lowest value observed among X-ray bursters (section~\ref{intro}).
\citet{Dohi+24} investigated the $\eta$ dependence on the equation of state (EOS) and NS masses in the range of $1.1M_{\odot}$ to $2.0M_{\odot}$. 
They found that compacted NS models tend to have lower values of $\eta$.\footnote{Note that their results were obtained for the case of H-rich accreted fuel, unlike the He-enhanced scenario we expected. However, because no significant correlation between the power-law index $\eta$ and the composition of the accreted fuel was found \citep{Lampe2016}, their conclusion holds.}
The value of $\eta = 0.84^{+0.02}_{-0.01}$ for SRGA J1444 does not match the model predictions for NS masses below $2.0M_{\odot}$, suggesting that it is a more massive NS.
A comparison of observed value $\eta$ in SRGA J1444 with theoretical models that encompass a broader range of masses above $2.0M_{\odot}$ could provide valuable insights for constraining the mass and the EOS of NSs.

The maximum observed value of the burst recurrence time can be used to constrain the composition of the accreted fuel.
Both the NinjaSat and IXPE observations indicate that the $\Delta t_{\rm rec}$--$F_{\rm per}$ relation with $\eta \sim 0.8$ remained valid up to at least $7.9$~hr.
This implies that during this period, H burned stably via the hot CNO cycle between each burst without depletion, followed by a mixed H/He burst (Case 1; \cite{Fujimoto1981}).
In this bursting regime, the accreted H is depleted in time of $t_{\rm CNO} = 9.8\ {\rm hr}\ (X/0.7)(Z_{\rm CNO}/0.02)^{-1}$ (\cite{Lampe2016}).
Using the maximum recurrence time of $\Delta t_{\rm rec} = 7.9$~hr observed in SRGA J1444, we obtain a constraint 7.9~hr $\leq (1+z)t_{\rm CNO}$, where $1+z$ represents the gravitational redshift at the photosphere. 
Assuming a solar hydrogen abundance of $X = 0.74$, along with a canonical NS with a mass $M_{\rm NS} = 1.4 M_{\odot}$ and the radius $R_{\rm NS} = 11.2$~km (giving $1 + z = 1.259$), we derive an upper limit for the CNO mass fraction as $Z_{\rm CNO} \leq 0.033$. 
Similarly, for a massive NS with $M_{\rm NS} = 2.0 M_{\odot}$ and the radius $R_{\rm NS} = 11.2$~km, the upper limit becomes $Z_{\rm CNO} \leq 0.038$.
Note that for the He-enhanced scenario with $X/Y = 1.5$ and $Z = 0.015$,  as inferred by \citet{Dohi2025PASJL}, $t_{\rm CNO}$ is estimated to be 11.0~hr, satisfying the constraint $7.9~\text{hr} < (1 + z) t_{\rm CNO}$ regardless of the NS mass.

This study demonstrated that CubeSat pointing observations can provide valuable astronomical X-ray data.
Even a compact detector, with an effective area of several tens of cm$^2$ onboard a CubeSat, can successfully observe both burst and persistent emissions. 
Given the rarity of X-ray bursts, which occur during less than 1\% of the total observation time \citep{Galloway2020}, CubeSats offer a complementary approach to large canonical observatories, providing long-term, flexible observations critical for detecting these transient events.

\vspace{-0.3cm}

\begin{ack}
This project was supported by JSPS KAKENHI (JP23KJ1964, JP17K18776, JP18H04584, JP20H04743, JP20H05648, JP21H01087, JP23K19056, JP24H00008, JP24K00673). 
T.T. was supported by the JSPS Research Fellowships for Young Scientists.
T.E. was supported by ``Extreme Natural Phenomena'' RIKEN Hakubi project. N.N. received support from the RIKEN Intensive Research Project (FY2024–2025). 


\end{ack}

\vspace{-0.3cm}

\bibliographystyle{apj}
\bibliography{reference}

\end{document}